%% file: ICMLT_24.tex
\documentclass[sigconf]{acmart}

\input{commands}

\AtBeginDocument{%
  \providecommand\BibTeX{{%
    \normalfont B\kern-0.5em{\scshape i\kern-0.25em b}\kern-0.8em\TeX}}}

\copyrightyear{2024} 
\acmYear{2024} 
\setcopyright{rightsretained} 
\acmConference[ICMLT 2024]{2024 9th International Conference on Machine Learning Technologies (ICMLT)}{May 24--26, 2024}{Oslo, Norway}
\acmBooktitle{2024 9th International Conference on Machine Learning Technologies (ICMLT) (ICMLT 2024), May 24--26, 2024, Oslo, Norway}
\acmDOI{10.1145/3674029.3674069}
\acmISBN{979-8-4007-1637-9/24/05}

\begin{document}

\title{Enhancing Cross-Market Recommendation System with Graph Isomorphism Networks: A Novel Approach to Personalized User Experience}

\author{Sümeyye Öztürk}
\email{ozturks20@itu.edu.tr}
\affiliation{%
  \institution{Istanbul Technical University}
  \city{Istanbul}
  \country{Turkey}
}

\author{Ahmed Burak Ercan}
\affiliation{%
  \institution{Istanbul Technical University}
  \city{Istanbul}
  \country{Turkey}}
\email{ercana20@itu.edu.tr}

\author{Resul Tugay}
\affiliation{%
  \institution{Gazi University}
  \city{Ankara}
  \country{Turkey}}
\email{resultugay@gazi.edu.tr}

\author{Şule Gündüz Öğüdücü}
\affiliation{%
  \institution{Istanbul Technical University}
  \city{Istanbul}
  \country{Turkey}}
\email{sgunduz@itu.edu.tr}

\renewcommand{\shortauthors}{S. Ozturk et al.}

\begin{abstract}
    In today's world of globalized commerce, cross-market recommendation systems (CMRs) are crucial for providing personalized user experiences across diverse market segments. However, traditional recommendation algorithms have difficulties dealing with market specificity and data sparsity, especially in new or emerging markets. In this paper, we propose the CrossGR model, which utilizes Graph Isomorphism Networks (GINs) to improve CMR systems. It outperforms existing benchmarks in NDCG@10 and HR@10 metrics, demonstrating its adaptability and accuracy in handling diverse market segments. The CrossGR model is adaptable and accurate, making it well-suited for handling the complexities of cross-market recommendation tasks. Its robustness is demonstrated by consistent performance across different evaluation timeframes, indicating its potential to cater to evolving market trends and user preferences. Our findings suggest that GINs represent a promising direction for CMRs, paving the way for more sophisticated, personalized, and context-aware recommendation systems in the dynamic landscape of global e-commerce.
\end{abstract}



\maketitle

\section{Introduction}

Recommendation systems play a crucial role in providing personalized user experiences across various fields like e-commerce, entertainment, and content platforms. Most of these systems provide recommendations exclusively for items within a specific or single domain. Yet, traditional methods encounter challenges in target markets where data is scarce or unavailable, along with substantial variations in user behaviors. To address these challenges, Cross-Domain Recommendation (CDR) and Cross-Market Recommendation (CMR) have emerged as promising approaches \cite{1}. Although both approaches aim to enhance recommendation performance by utilizing information from different domains or markets, they address distinct challenges and find applications in different scenarios.

CDR is to make recommendations across domains or markets. It is concerned with providing suggestions in a target domain based on knowledge obtained from a source domain \cite{2}. For example, knowledge about a user's book purchases can also be used to offer suggestions on some other seemingly irrelevant topics, like film or music. This approach improves the user experience because there may be dependencies and correlations between user preferences in different domains. Consider a multinational e-commerce company operating across many countries in Europe and is looking to expand into a new market, like Spain, where there is limited local data available. The entity can improve the quality of its recommendation algorithms in Spain by using strong data from mature markets. This approach not only customizes the shopping experience for the Spanish audience but also opens up the door for the company's growth in the region. 

CMR is the choice, for this research paper for a variety of reasons. Firstly, it allows businesses to explore markets and seize opportunities by using data from established markets \cite{1}. This means that when faced with data in a new market personalized recommendations can still be offered to users enhancing their experience and boosting engagement. Secondly, CMR tackles the challenge of data in target markets by utilizing data from source markets that have diverse interactions. This significantly improves the performance of recommendations in the target market \cite{3}. Lastly, it facilitates the discovery of user preferences and behaviors that may vary across markets \cite{4}. By using insights from source markets, the system can adapt to users' distinct characteristics and preferences, in the target market.

Traditional recommendation methods, such as collaborative filtering \cite{5} and content-based filtering \cite{6}, have been widely used for making recommendations in one market. However, these methods often rely on user-item interactions or item attributes within a specific market, making them less effective in understanding the complexities of CMRs. The limitations of traditional methods in the context of CMR can be attributed to several factors. Firstly, the lack of data compatibility and heterogeneity across markets presents challenges in using user-item interactions or item attributes from one market to another. Additionally, the differences in user preferences, market dynamics, and product characteristics across markets further complicate the recommendation task \cite{7}.

These conventional approaches depend on user-item interactions or item attributes within a market, which restricts their usefulness in market activities. Therefore, to provide recommendations that apply across markets, it is necessary to employ advanced strategies capable of navigating the intricacies of the global market \cite{1}. Researchers have turned to Graph Neural Networks (GNNs) to capture relationships and dependencies in graph data to tackle these challenges \cite{9}. \eat{GNNs have gained popularity in recommender systems and have demonstrated exceptional capabilities in acquiring meaningful user-item representations from their interactions \cite{10}.} \revise{GNNs have gained popularity in recommender systems and have demonstrated exceptional capabilities in creating user-item representations. Unlike traditional methods, GNNs can incorporate neighborhood information and relational dependencies among items and users, providing a more comprehensive and contextually enriched representation \cite{10}.}

Using GNNs in CMR systems can offer more personalized and on-demand suggestions \cite{11}. Compared to traditional methods, GNNs offer great performance in CMRs due to their ability to manage sparse data and also social information. \eat{GNN-based models further include \warn{contrastive learning} to help the task of the recommender system \cite{12}.} \revise{GNN-based models utilize high-order connectivity, which refers to the capability of these models to capture relationships beyond immediate neighbors in a graph structure \cite{13}.} Traditional GNNs typically consider only first-order connections, i.e., direct neighbors of a node \cite{14}. However, high-order connectivity extends this concept to include connections up to k-th order, where k is a parameter that determines the depth of exploration in the graph. By incorporating high-order connectivity, GNNs can effectively capture more complex patterns and dependencies in the data. This allows GNN-based models to leverage not only direct interactions but also indirect relationships between nodes, enabling a more comprehensive understanding of the underlying graph structure. In the context of collaborative filtering and recommendation systems, high-order connectivity enables GNNs to extract valuable information from distant nodes in the graph, leading to more accurate and personalized suggestions for users \cite{15}. Furthermore, high-order connectivity enhances the ability of GNNs to handle sparse data by using information from nodes that are not directly connected but are reachable through intermediate nodes. This is particularly beneficial in scenarios where data sparsity is a common challenge, as it enables GNNs to exploit the full potential of the available information in the graph \cite{16}.

\section{Literature Review}

Cross-market recommendation has made significant progress in recent years, with researchers exploring traditional methods as well as GNNs to tackle the challenges posed by data sparsity and cold-start issues in target markets \cite{17}. \eat{This literature review discusses several important papers that have contributed to CMRs.}

By combining deep graph recommendation models like GNNs and conventional models like ItemCF \cite{18} and UserCF \cite{19}, CMR leverages colloborative signals and similarities across marketplaces. The efficacy of such an approach, employing a combination of models including DeepWalk \cite{20} and LightGBM \cite{21}, has been notably shown on the XMRec dataset \cite{22}.

Creating generalized representations for CMR has centered on integrating self-supervised graph learning. SGLCMR, models market-specific aspects by creating single-market graphs and market\-generalized features by creating cross-market graphs \cite{23}. Extensive studies on real-world datasets have demonstrated the superiority of this methodology over state-of-the-art CMR methods, since of its utilization of Light Graph Convolution (LGC) \cite{24} and self-supervised graph learning.

With models like Bert4CMR, the attention mechanism has also been investigated in CMR with the goal of enhancing recommendation performance across markets \cite{25}. This model uses market embedding to simulate each market's bias and minimize mutual interference between parallel markets, demonstrating cutting-edge outcomes from comprehensive testing.

Graph Attention Networks (GATs) \cite{26} was a significant breakthrough that addressed the shortcomings of previous graph convolution approaches and allowed for more complex interactions between nodes in a variety of applications, including protein-protein interaction datasets and citation networks. This breakthrough was extended by the HGNNDR model \cite{27}, which improved node feature representation in situations with sparse network connections by using multi-order topology information. This improved the quality of recommendations. Similarly, to address the issues of data sparsity and cold-start in item recommendation, the TransD-KGAT method \cite{28} used knowledge graphs, highlighting the effect of GNN-based approaches in recommendation tasks.

Models such as GraphRec \cite{29} and SR-GNN \cite{30} became specialized frameworks concentrating on social and session-based recommendations. These models demonstrate the adaptability of GNNs in capturing complicated relationships between users and items in a variety of recommendation scenarios, offering complete representations of user-item relationships and enhancing recommendation accuracy. They also complement the enhanced collaborative filtering recommendation model \cite{31}.

With the introduction of FOREC, the field also witnessed notable advancements in cross-market product recommendation, where it demonstrated its supremacy on the XMarket dataset by utilizing data from both source and target markets \cite{1}. In addition, the Heterogeneous Graph Attention Network (HAN) is introduced, which extends GNNs to heterogeneous graphs and shows that it can manage diverse data difficulties and achive better performance on real-world heterogeneous grahps \cite{32}.

\eat{\warn{Apart from the previously mentioned papers, there exist other significant developments in cross-market recommendation through the use of graph neural networks.} A recommendation algorithm based on graph factorization machines that incorporates an attention mechanism and graph construction is presented \cite{33}. They employ graph neural networks to explicitly enhance recommendation performance by encoding collaboration signals. Neural Graph for Personalized Tag Recommendation (NGTR) model \cite{34} leverages graph neural networks to learn tag representation and increase tag recommendation accuracy.}

\revise{Recent advancements in CMR systems have showcased the versatility and efficacy of GNNs. One notable example includes a recommendation algorithm based on graph factorization machines that incorporates an attention mechanism and graph construction \cite{33}. This approach utilizes GNNs to notably enhance recommendation performance by encoding collaboration signals. Similarly, the Neural Graph for Personalized Tag Recommendation (NGTR) model \cite{34} employs GNNs to improve tag representation learning, thereby increasing the accuracy of tag recommendations.}

\begin{figure*}[!ht] 
\centering
\includegraphics[width=\textwidth]{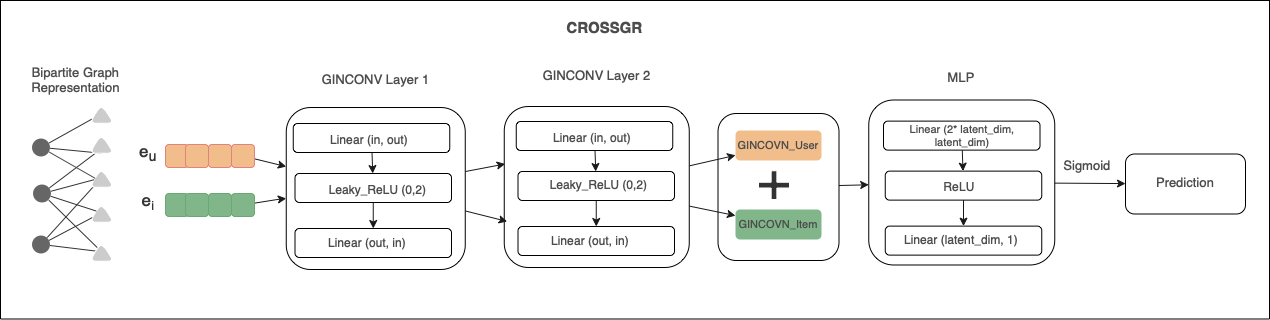}
\eat{\caption{Overview of the CrossGR framework for enhancing cross-market recommendations. The model begins with a bipartite graph representation that captures the interactions between users and items. It then utilizes two GINConv layers to transform these interactions into an embedded space. The combined user and item embeddings undergo further transformation through an MLP to predict the interaction scores. A sigmoid function is applied to convert the scores into probabilities, providing the final recommendation ratings.}}
\caption{Overview of the CrossGR framework for enhancing cross-market recommendations. Input of the model is a bipartite graph that represents interactions between users and items. The model utilizes two GINConv layers to transform these interactions into an embedding space. Then concationation of these embbeddings is passed through an MLP layer to predict the interaction scores. Lastly, a sigmoid function is applied to convert the scores into probabilities, providing the final recommendation prediction.}
\label{fig:crossgr_framework}
\end{figure*}

\section{METHODOLOGY}

To address CMR challenges, we propose "CrossGR," a new method utilizing GIN for their ability to analyze complex user-item graphs, enhancing CMR effectiveness. GINs, introduced by Xu et al. (2018) \cite{35}, matches the Weisfeiler-Lehman test in graph analysis prowess, offering a nuanced understanding of graph topology, which marks a significant advancement over prior graph neural network models. This approach allows "CrossGR" to more effectively harness the inherent topological features of graphs for improved recommendation accuracy.

The motivation behind GIN for our CMR task arises from its unique aggregation function that updates node representations by summing the features of neighboring nodes, including the node itself. This approach ensures that the network can distinguish between different graph structures, a property known as injectivity, which is critical when dealing with diverse and sparse data inherent in cross-market scenarios. The steps applied in our methodology, from data representation to model optimization, are visually summarized in Figure \ref{fig:crossgr_framework}, which complements the detailed explanations provided in the following subsections.

\subsection{Data Representation}

\subsubsection{Notation}

In the construction of our graph-based data representation for the CMR system, we encapsulate the user-item interactions within graph $G$, composed of a set of nodes $V$ and a set of edges $E$. The nodes in $V$ represent both users and items in the dataset, drafted as sets $U$ and $I$ respectively, reflecting the dual nature of our marketplaces. Interactions between users and items, such as ratings or purchases, form the edges in $E$, bridging the two distinct sets within our graph. To quantify the interactions encapsulated by $E$, we introduce an adjacency matrix $A$, where the weight of each edge $A_{ij}$ signifies the strength and nature of the interaction between user $i$ and item $j$. This setup inherently forms a bipartite graph, capturing the distinct user-item relationships within our system.

To ensure that our model can learn from these interactions without bias toward nodes with higher degrees of connectivity, we normalize the adjacency matrix using the diagonal degree matrix $D$, where each diagonal element $D_{ii}$ corresponds to the sum of weights connected to node $i$. The normalization process, yielding $\tilde{A} = D^{-1}A$, ensures that the influence of each node is adjusted by its connectivity, facilitating a balanced and effective learning process.

\subsection{Graph Isomorphism Network Architecture}

Our CrossGr model employs embeddings for initially representing users and items, mapping their features into a latent space. This foundational step is crucial for capturing the essence of user-item interactions. Subsequently, through the application of \warn{GINConv} layers, we refine these embeddings by aggregating and transforming features from their neighborhoods. This process enhances the model's ability to accurately reflect the complex dynamics of interactions.

The GIN leverages a message-passing framework that iteratively updates node representations by aggregating features from their neighborhoods, thus encapsulating local and global market structures. The model's forward pass involves a series of convolutional operations, each comprising two key components: feature transformation and neighborhood aggregation. Specifically, the GINConv layer employs an "add" aggregation function – a choice motivated by its effectiveness in preventing the over-smoothing of node features, a common challenge in deep GNNs where node representations can become indistinguishable.

The embeddings are updated through the forward pass of the network, where GINConv aggregates features from the nodes' neighborhoods, refining the representations based on local and global structural information. The convolutional layers in our model are designed to enhance the expressiveness of node embeddings. They employ learnable linear transformations followed by a non-linear activation function, ReLU, \eat{\warn{to ensure the injective nature of the aggregation process.}} The layers are parameterized to allow for the flexibility needed to capture the diverse interaction patterns found in cross-market data. The process can be mathematically expressed as:
\begin{equation}
  H^{(k+1)} = MLP^{(k)} \left( (1 + \epsilon^{(k)}) \cdot H^{(k)} + \sum_{u \in N(v)} H_u^{(k)} \right)
\end{equation}

Where $H^{(k)}$ represents the node features at the $k^{th}$ iteration, $N(v)$ denotes the set of neighbors of node $v$, and $\epsilon$ is a learnable parameter that modulates the contribution of the node's features in the aggregation process. An MLP (Multi-Layer Perceptron) allows for a more profound combination of features before updating the node representation.

In practice, the embeddings are passed through a sequence of ReLU-activated GINConv layers, followed by a final MLP that combines user and item embeddings to predict the interaction score. The MLP consists of a linear layer that doubles the embedding size, a ReLU activation, and another linear layer that projects the combined embeddings down to a single score. This score is then passed through a sigmoid function to yield a probability, representing the likelihood of interaction between the user and the item. The process can be mathematically expressed as follows:

\begin{equation}
     H^{(k+1)} = \text{ReLU}\left(\text{GINConv}\left(H^{(k)}, A\right) + \epsilon^{(k)} \cdot H^{(k)}\right)
\end{equation}

Where \( H^{(k)} \) denotes the hidden representations of nodes at the \( k \)-th layer, \( A \) is the adjacency matrix, and \( \epsilon^{(k)} \) is a learnable parameter that allows for distinguishing the node's features from its neighbors. The GINConv operation performs a sum aggregation of neighbor features followed by a linear transformation.

\subsection{Model Optimization and Hyperparameters}

The model concatenates user and item embeddings to predict interaction scores via a Multi-Layer Perceptron (MLP). We employ Binary Cross-Entropy (BCE) as our loss function:

\begin{equation}
    \mathcal{L} = -\frac{1}{N} \sum_{(u,i) \in D} \left( y_{ui} \log(\sigma(\hat{y}_{ui})) + (1 - y_{ui}) \log(1 - \sigma(\hat{y}_{ui})) \right)
\end{equation}

where \( D \) is the set of user-item pairs, \( y_{ui} \) is the true label, \( \hat{y}_{ui} \) is the predicted score, and \( \sigma \) denotes the sigmoid function. The loss is minimized using the Adam optimizer with a learning rate \( \alpha \), and other hyperparameters such as the mini-batch size \( B \), the weight decay coefficient \( \lambda \), and dropout rate \( p \) to prevent overfitting. The specific values for these hyperparameters are empirically determined through a grid search on the validation set.

For model optimization, we utilize the Adam optimizer with a learning rate specified by adam\_lr. The choice of Adam is due to its adaptive learning rate capabilities, which are particularly beneficial in the sparse and skewed distribution of CMR datasets. This process iteratively adjusts the embeddings and the weights of the GINConv and MLP layers to enhance the model's predictive performance on the training data.

\section{EXPERIMENTS}

In this section, we evaluate the performance of our proposed CrossGr model through a series of  experiments. We aim to demonstrate CrossGr's effectiveness in enhancing cross-market recommendation systems compared to existing models. To prepare for our experimental analysis, we have divided this section into several key subsections. Firstly, we describe the datasets used in our study, highlighting their relevance and the challenges they pose for cross-market recommendation systems. Then, we provide details about the experimental setup, including parameter settings and the reasons for selecting certain evaluation metrics. Finally, we discuss the implications of our findings, providing insights into the potential impact of CrossGr on the field of recommendation systems. This structured approach ensures a comprehensive evaluation of our model's capabilities and comparative effectiveness.

\subsection{Data \& Data Analysis}

The Dataset provided by the competition organizer \cite{Dataset} contains user interactions from five distinct markets, detailed in Table \ref{tab:1}. This table presents a breakdown of the datasets into three source markets (s1, s2, and s3) with a substantial number of users and interactions, highlighting the breadth of consumer data available for analysis. Additionally, two target markets (t1 and t2) are included, showcasing a more focused set of interactions that will be crucial for testing the transferability of our recommendation system. The variation in the number of users and interactions across these markets underscores the need for a recommendation system that can adapt to different levels of market data density and user-item interaction frequency.

\begin{table}[h]
\centering
\caption{General Description of Datasets}
\label{tab:1}
\begin{tabular}{crrr}
\hline
\textbf{Dataset} & \textbf{\# Users} & \textbf{\# Items} & \textbf{\# Interactions} \\
\hline
s1 & 6466 & 9762 & 77173 \\
s2 & 7109 & 2198 & 48302 \\
s3 & 3328 & 1245 & 23367 \\
t1 & 2697 & 1357 & 19615 \\
t2 & 5482 & 2917 & 41226 \\
\hline
\end{tabular}
\end{table}

\subparagraph{\textbf{Dataset Structure: }}

\begin{table}[t]
\centering
\caption{Summary of Overlap Items}
\label{tab:2}
\begin{tabular}{lrrrrll}
\toprule
Market &  Total &   s1 &  s2 &  s3 &  t1 &  t2 \\
\midrule
    t1 &   1357 &  796 & 863 & 712 &   - & 530 \\
    t2 &   2917 & 1834 & 789 & 548 & 530 &   - \\
\bottomrule
\end{tabular}
\end{table}

For an effective CMR system, understanding the shared item count across markets is essential. Our analysis, depicted in Table \ref{tab:2}, focuses on the item overlap between different markets. Importantly, it should be noted that there are no common users between these markets, underscoring the distinct nature of each market's user base and the unique challenge it presents in CMRs. This information can potentially assist in closing the gap between user preferences and expectations across different markets.

\subsubsection{Data Characteristics}

The following highlights were obtained from the dataset:
\begin{itemize}
    \item The market ratings are mostly between 4 to 5 with an average of 4.6, indicating the feasibility of direct knowledge transfer based on ratings.
    \item The users in each market are distinct and non-overlapping to prevent overfitting of specific user behaviors.
    \item The items, on the other hand, exhibit a pronounced overlap, especially in the target markets. This reinforces the significance of using item information across markets to improve the recommendation system's accuracy.
\end{itemize}

\subsection{Evaluation Metrics}

In order to evaluate the performance of our recommendation system simply and accurately, we utilize a strict evaluation methodology that attaches to the highest standards in the field. \eat{\warn{Our approach is based on the XMRec dataset}, which provides a wealth of cross-market user-item interaction data, ensuring our results are based on real-world e-commerce scenarios.}\revise{This methodology utilizes the XMRec dataset, a comprehensive collection of cross-market user-item interaction data. The real-world e-commerce scenarios provided by this dataset serve as a robust basis for assessing our system's effectiveness.}

We measure the effectiveness of our model using Hit Rate (HR) and Normalized Discounted Cumulative Gain (NDCG), with a specific focus on the top-10 recommendations. HR directly measures whether the actual items of interest to users are captured within the top-N recommendations. Formally, HR is defined as:

\begin{equation}
HR = \frac{1}{|U|} \sum_{u \in U} \delta(p_u \leq topN)
\end{equation}

where \(\delta\) is the indicator function, \(p_u\) denotes the rank position of the test item for user \(u\), and \(U\) represents the user base.

\eat{NDCG \warn{introduces a progressive perspective} by accounting for the rank order of the recommendations, attaching greater value to hits that occur higher in the list. This metric is especially relevant when the rank of a recommendation is as critical as its presence. NDCG is defined as:}

\revise{NDCG emphasizes the importance of the ranking order in the recommendation list by assigning more value to highly placed recommendations. This approach ensures that recommendations appearing earlier, and thus more likely to be seen by users, are given priority in the evaluation. Such a metric becomes crucial in scenarios where the position of a recommendation is as important as its presence. NDCG is defined as:}

\begin{equation}
NDCG = \frac{1}{|U|} \sum_{u \in U} \frac{\log_2(1 + \delta(p_u))}{\log_2(p_u + 1)}
\end{equation}

These metrics, HR and NDCG, serve as the cornerstone of our evaluation, providing a dual-lens view of our model's ability to predict preferences accurately and rank them in a manner that aligns with user expectations and behaviors. Through these metrics, we aim to present a comprehensive analysis of our model's performance on the XMRec dataset, thereby substantiating the effectiveness of our CMR approach.

\subsection{Baselines}

To establish the comparative efficacy of our proposed Graph Isomorphism Network (GIN) for CMRs, we compare its performance with several well-established baselines. These models, recognized for their proficiency in conventional single-market recommendation tasks, serve as a benchmark to highlight the advancements our GIN model brings to the cross-market domain. The baselines included in our evaluation are:

\begin{itemize}
    \item GMF++ \cite{36}: An enhanced version of the Generalized Matrix Factorization (GMF) model, GMF++ extends the traditional matrix factorization approach by integrating a linear kernel to capture latent features. This model excels in distilling user-item interactions into a latent space where recommendations are derived from the linear interactions of these features.
    \item MLP \cite{36}: Using a nonlinear kernel, the Multilayer Perceptron (MLP) delves into the complexity of user-item interaction data. It employs a deep architecture to model intricate and high-level feature interactions, which are not directly observable through linear methods alone. This allows MLP to capture the slight patterns of user preferences and item attributes.
    \item NMF \cite{36}: The Neural Matrix Factorization (NMF) model is a hybrid approach that combines the strengths of GMF and MLP. It takes advantage of both the linearity of GMF and the nonlinearity of MLP to uncover a rich hierarchy of features and interactions. With this comprehensive model for recommendation systems, NMF offers a powerful solution that can provide valuable insights.
    \item ItemCF \cite{18}: ItemCf algorithm recommends items to users based on their past interests. It does this by finding items that are similar to ones that the user has previously shown interest in. The goal is to provide more personalized recommendations that reflect the user's own interests.
    \item UserCF \cite{19}: UserCF algorithm recommends items based on the interests of other users who share similar preferences. It focuses on groups with similar interests and recommends popular items within those groups. This approach results in more social recommendations that reflect the popularity of items among users with similar interests.
\end{itemize}

\subsection{Hyperparameter Configuration}

In our search for a fair and strict comparison, we carefully optimized the hyperparameters of our GIN based recommendation system. This optimization is crucial to ensure that our model operates at its full potential when addressing the unique challenges posed by CMR tasks.

The model's parameters were  initialized from a Gaussian distribution \( \mathcal{N}(0, 0.1^2) \), setting the stage for a diverse exploration of the solution space during the training process. We employed the Adam optimizer, celebrated for its adaptive learning rate capabilities, which is particularly beneficial given cross-market data's diverse and sparse nature. The learning rate was set to a value of \( 0.01 \), chosen to balance convergence speed and stability.

The batch size, a critical parameter influencing the gradient estimation, was set to \( 1024 \), providing a representative sample of user-item interactions while maintaining computational efficiency. Regularization is critical to preventing overfitting, particularly in models with a high capacity such as ours; thus, we applied a weight decay, denoted by \( \lambda_2 \), of \( 1 \times 10^{-7} \).

\eat{The GIN model's capacity to encode user and item features into a latent space was determined by the latent dimension size, set according to the \texttt{\warn{--latent\_dim}} argument provided during model configuration. To ensure that our model learns to differentiate between positive and negative interactions, we specified the number of negative samples per positive interaction as determined by the \texttt{--num\_negative} argument, which was set based on preliminary experiments.}

\eat{The GIN model's capacity to encode user and item features into a latent space was determined by the latent dimension size, set at \( 8 \) according to the \texttt{--latent\-dim} argument provided during model configuration. To ensure that our model learns to differentiate between positive and negative interactions, we specified the number of negative samples per positive interaction as \( 4 \) determined by the \texttt{--num\_negative} argument, which was set based on preliminary experiments.}

\revise{The GIN model's capacity to encode user and item features into a latent space was determined by the latent dimension size, set at \( 8 \). To ensure that our model learns to differentiate between positive and negative interactions, we specified the number of negative samples per positive interaction as \( 4 \).}

The model's architecture includes user and item embeddings, which remain flexible to adjust various configurations, and we allow for the possibility of pre-trained embeddings to be incorporated through the \texttt{embedding\_user} and \texttt{embedding\_item} settings. We also enable the saving of the trained model via the \texttt{save\_trained} parameter, facilitating the utilization of the model without retraining.

The count of users and items, which limits the size of the embedding matrices, was dynamically set based on the highest index values observed in the data, ensuring that our model's capacity aligns with the dataset's scope.

\eat{
Lastly, the utilization of CUDA was determined based on system availability, allowing for accelerated computation if a compatible GPU device is present. This is reflected in the \texttt{use\_cuda} parameter, which, along with the \texttt{device\_id}, ensures that our model runs on the appropriate hardware, thus optimizing training efficiency.}

Through these carefully tuned hyperparameters, our model is poised to learn from the intricate patterns of cross-market interactions and generalize these learnings to new, unseen market data, pushing the boundaries of what is possible in CMRs.

The model is trained over a series of epochs, with the best model state saved based on the validation performance, specifically monitoring the Normalized Discounted Cumulative Gain (NDCG@10) and Recall@10. These metrics guide the training process and facilitate early stopping to prevent overfitting.

\subsection{Results and Discussion}

\begin{table}[htbp]
\centering
\caption{Comparative Performance Analysis of Recommendation Algorithms on Cross-Market Datasets Highlighting Top Achievers in NDCG@10 and HR@10 Metrics}
\label{tab:3}
\begin{tabular}{@{} lccccc @{}}
\toprule
Method & \multicolumn{2}{c}{t1} & \multicolumn{2}{c}{t2} \\
\cmidrule(r){2-3} \cmidrule(r){4-5}
& NDCG@10 & HR@10 & NDCG@10 & HR@10 \\
\midrule
GMF++ & 0.3343 & 0.5190 & 0.3397 & 0.4574 \\
MLP & 0.3245 & 0.5113 & 0.3309 & 0.5222 \\
NMF & 0.5456 & 0.6078 & 0.4898 & 0.5932 \\
ItemCF & 0.4838 & 0.5622 & 0.4132  & 0.5056  \\
UserCF & 0.5267 & 0.5897 & 0.4472 & 0.5643 \\
CrossGR & 0.6524 & 0.7609 & 0.6113 & 0.7206 \\
\end{tabular}
\end{table}

In the process of evaluating various recommendation algorithms across different market datasets, our study focused on two key performance metrics: NDCG@10 and HR@10. These metrics were selected for their ability to effectively measure the accuracy and relevance of the recommendations, as well as the models' capacity to prioritize the most relevant items for each user. We evaluated the performance of several algorithms, including GMF++, MLP, NMF, ItemCF, UserCF, and our proposed method, CrossGR, over two distinct evaluation periods, t1 and t2, to measure both their immediate efficacy and their stability over time. Table \ref{tab:3} illustrates the performance of these models. 

From the results (Table \ref{tab:3}), CrossGR emerges as the superior model, achieving the highest scores in both NDCG@10 and HR@10 across both evaluation periods. Specifically, in the t1 market, CrossGR recorded an NDCG@10 of 0.6524 and an HR@10 of 0.7609. Its performance in the t2 market, with an NDCG@10 of 0.6113 and an HR@10 of 0.7206, underscores its consistency and reliability over time, surpassing the other models evaluated.
Comparatively, traditional models like GMF++, MLP, NMF and even collaborative filtering methods such as ItemCF and UserCF, displayed lower performance metrics. For instance, NMF, the closest competitor, achieved an NDCG@10 of 0.5456 and an HR@10 of 0.6078 during t1, and an NDCG@10 of 0.4898 and an HR@10 of 0.5932 in t2, significantly trailing behind CrossGR.

Our proposed method, CrossGR, utilizes a Graph Isomorphism Network model that incorporates a novel approach to encoding user and item features into a high-dimensional latent space. This approach is crucial for capturing the complex and non-linear relationships inherent in user-item interactions, particularly in cross-market scenarios where user preferences and item characteristics may vary significantly across datasets.

We conducted  hyperparameter optimization to fine-tune the GIN model, providing it operates optimally under the unique constraints of cross-market recommendations. The model's parameters were initialized from a Gaussian distribution, with the Adam optimizer selected for its adaptive learning rate properties. This was particularly useful for managing the sparse and diverse data encountered in cross-market recommendation systems.

The GIN model's architecture is designed to capture the intricate patterns of user-item interactions. By employing a Multi-Layer Perceptron (MLP) to predict interaction scores, the model can leverage deep learning to enhance the accuracy of its recommendations. Binary cross-entropy serves as the loss function, guiding the model towards more precise predictions. We also implemented regularization techniques, such as weight decay and dropout, to mitigate the risk of overfitting, ensuring that the model remains generalizable and robust across different datasets.

The evaluation methodology was designed to reflect real-world e-commerce scenarios accurately. We utilized the XMRec dataset, renowned for its comprehensive cross-market user-item interaction data. The use of NDCG@10 and HR@10 metrics provided a comprehensive view of each model's performance, evaluating not only the accuracy of the recommendations but also their ranking in alignment with user preferences.

CrossGR was benchmarked against several well-established models to highlight its relative performance advantages. This comparative analysis underscores the significant improvements our model introduces, particularly in the context of cross-market recommendations, where traditional models often struggle to account for the variability and diversity of user preferences and item attributes across different markets.

\section{Conclusions}

In this work, we have presented a novel approach to CMRs by employing Graph Isomorphism Networks (GINs), which have demonstrated capability in capturing complex user-item interaction patterns within diverse market scenarios. Our approach, called CrossGR, has shown significant improvement over traditional recommendation systems in terms of NDCG@10 and HR@10 metrics. CrossGR is adaptable and accurate, which makes it a potential tool for improving user engagement and satisfaction across different market segments. Our research suggests that GINs can effectively handle the complexities of cross-market data, and provide a scalable solution for personalized recommendation services in a globally interconnected marketplace.

\eat{\warn{
In the future, we plan to conduct more research to improve the model's architecture, scalability, and extend its applicability to other recommendation systems}. Our research has shown a promising direction towards more advanced, context-aware, and user-centric recommendation strategies. Ultimately, this will enhance consumers' experiences in the digital marketplace.}

\eat{In future work, we intend to explore the integration of heterogeneous market data within cross-market recommendation systems. This research will aim to enhance the model’s architecture and scalability by devising methods for the effective combination of datasets from varied markets, thus extending its applicability. Our initial findings suggest a viable path towards developing more sophisticated, context-aware, and user-centric recommendation algorithms. By investigating the interplay between different market dynamics, we plan to generate insights that will refine our approach to recommendation systems. Such improvements are anticipated to elevate the consumer experience in the digital marketplace, by providing recommendations that are more accurately aligned with individual user preferences and behaviors across diverse market contexts.}

\revise{In future work, we intend to explore the integration of heterogeneous market data within cross-market recommendation systems. By investigating the interplay between different market dynamics, we plan to generate insights that will refine our approach to recommendation systems.}

\bibliographystyle{ACM-Reference-Format}
\bibliography{sample-base}

\end{document}

%% file: commands.tex
\usepackage{amssymb}
\usepackage{latexsym}
\usepackage{amsfonts}
\usepackage{amsmath}
\usepackage{color}
\usepackage{xspace}
\usepackage{graphicx}
\usepackage{cleveref}
\usepackage{subfigure}
\usepackage{balance}
\usepackage{hhline}
\usepackage{float}
\usepackage{xcolor}
\usepackage{tabu}

\newcommand{\revise}[1]{{\color{black}{#1}}}

\newcommand{\warn}[1]{{\color{black}{#1}}}

\newcommand{\eat}[1]{}